# Efficient Transport Logistics

## An Approach for Urban Freight Transport in Austria

Verena Brandstätter, Cristina Olaverri-Monreal*
Chair Sustainable Transport Logistics 4.0
Johannes Kepler University
Linz, Austria
verena-brandstaetter@gmx.at, cristina.olaverri-monreal@jku.at

*Abstract*— **To alleviate traffic congestion that results from the growth of e-commerce we propose an approach in the city of Linz, Austria by relying on shared distribution centers from different companies. We develop two algorithms to find out the optimal location for the hubs and calculate the shortest path between locations. Results showed that in an urban environment, the implementation of hubs results in a reduction of the number of delivery vehicles. It reduces driving distances from hub to the customers, and also benefits the drivers that need to return home every day.**

**Keyword-components: transport logistics; routing; sustainability; traffic**

## I. INTRODUCTION

According to the United Nations the transport sector is responsible for a quarter of the energy-related greenhouse gas emissions (GHG) worldwide [1]. This is due to a traffic increase resulting from the growth of e-commerce. The future promises only more vans and packages. Since in this sector the emissions increase at a faster rate than in other sectors, it is crucial to take measures to reduce global temperature increase.

To reduce congestion and the negative impact of transport on the environment, while also contributing to a decrease in cities' agglomerations in the last kilometers before the delivery, a common and efficient approach is to tackle the last 2 to 5 kilometers to the customer. This can be done by creating micro-hubs in the city center, storage facilities from which packages are distributed in the surrounding area. By relying on cooperative approaches between different logistic companies, such as sharing and pooling resources for bundling deliveries in the same zone, an additional environmental benefit can be reached. The goal is to treat all the urban freight transport stakeholders, material elements and activities as a part of an integrated logistic system instead of each one of them individually [2].

Shared distribution centers or hubs have been suggested in several works to be used by many different companies as part of a common, open and interconnected logistic network [3].

On a more practical side, last mile delivery approaches based on automation and intelligent boxes have been presented in [4].

Another practical example can be found in [5] which is a case study of the eastern Canada road network. The authors compared the conventional logistic system (a), with a system that relied on physical internet (b) and with a hybrid logistic system, similar to the previous one but in which the containers were not loaded and unloaded at every hub *(c)*. Results showed the gains of *(b)* in terms of short driving distances and environmental benefits.

In [6] the hubs were replaced with parcel lockers so that both customers as well as carriers could pick-up their packages. The tests with small and large data sets showed a good performance of the approach.

In the Netherlands urban consolidation centers were used to optimize the deliveries of small and independent retailers. Nowadays they are used for customers [7].

In Austria several projects focus on reducing the number of vehicles in the city. For example by establishing a city hub to deliver groceries by bike or by bundling parcels [8].

In this work we propose an approach with data from the city of Linz, Austria. We determine the best location for shared distribution centers from different companies that bundle their deliveries to reduce empty mileage, maximize their storage capacity and achieve the shortest routes to reduce driving and delivery time.

## II. SYSTEM REQUIREMENTS

To develop our approach, we rely on several assumptions to simplify the model and to make it feasible in real life scenarios. To this end, we develop two algorithms that are able to:

- find out the optimal location for the hubs and
- calculate the shortest path between two specific locations

### A. Customer Location Definition

Not every apartment and house in the city is considered as a potential customer. However, pick-up stations from the delivery services such as DPD, Hermes and UPS as well as parcel lockers from the post are always considered customers. Pick-up stations are often included in a shop where the package addressees can get their parcels. According to the size of the shop, the capacity is estimated between 20 and 50 parcels per shop. For the parcel lockers belonging to the Austrian post the capacity is 150 packages.

*B. Zones*

Since it would not be possible to build one hub with enough capacity for the whole city of Linz, we divided the city into three zones. Each zone was assigned one hub from which the packages could be delivered to all customers within the same zone. The division was calculated based on the density of the population, the ability to separate it from the adjacent zone and on overall traffic volume. Fig. 1 depicts the established zones.

- Zone 1 covers the whole area of Linz that is north of the Danube. In this area there are often traffic jams across the bridges. And therefore, to safe time, the amount of trucks that cross the bridges should be minimized. In this zone, there are 14 existing pick-up stations with a storage capacity between 20 and 50 packages and one parcel locker from the Post with a capacity for 150 parcels being their location between 48.310376 minutes latitude and 14.291251 longitude.

- Zone 2 is the area south of the Danube River until the line of the streets B139 – Unionstraße – Wiener Straße – Raimundstraße – Nebingerstraße. It includes the inner city and is quite small compared to the other two. However, the density of the pick-up stations and parcel lockers is quite high. It contains 20 pick-up stations (between 20 and 50 packages) and 3 parcel lockers (capacity 150), placed between 48.279143 latitude and 14.308681 longitude.

- Zone 3 is the southernmost area. The density of the customers is quite low and some areas are not covered at all. That means people have to drive long distances to get their deliveries. Hence, the storage capacity of the existing pick-up stations is between 10 and 35 packages. In this zone there are 12 pick-up stations and 3 parcel lockers (capacity 150) within the following coordinates: 48.246356 latitude and 14.3026191 longitude.

### III. HUB DEFINITION AND LOCATION

Once the locations of the customers are known, the next question is where to locate the hub. To solve the question, we used the Steiner-Weber-Model. Within this model every place in a certain defined zone, for example Linz, is a potential location for the hub. The Steiner-Weber-Model has several disadvantages:

- the demand has to be known in advance;
- costs are not considered;
- satisfying the demand with one central facility is not always possible.
- the Euclidian distance used in the model can differ a lot from the real driving distance between two locations

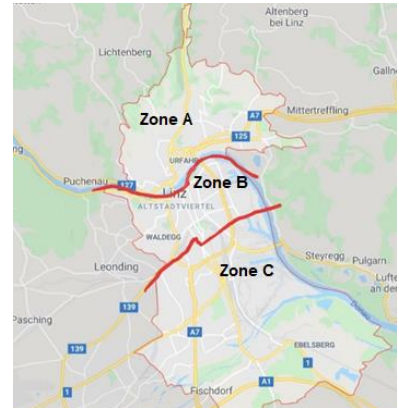

Figure 1. Established zones in Linz to implement the corresponding goods distribution centers

To minimize the disadvantages of the Steiner-Weber-Modell we proceed as follows addressing each of the issues above:

- The capacities are set equal to the potential demand.
- An algorithm is implemented to calculate driving distances.
- Costs are not so important in this model because there are no tolls or congestion charges in Linz. Hence, the only transportation cost is the gas that is used. Therefore, minimizing driving distance also means the minimizing costs.
- In order to have enough capacity for all packages Linz is divided into three zones specifying for each one center of distribution.

An algorithm is used to calculate the optimal location for the hubs. It considers the demand of the customers and aims at minimizing the distance from the customers to the hub. The level of accuracy of the solution determines the calculation time. We set in this work an accuracy of 0.00001 decimal degree of latitude and longitude.

### IV. HUB LOCATION ALGORITHM IMPLEMENTATION

To calculate the optimal location of the hub the algorithm needs the following input: 1) the number of customers, so that the correspondent number of rows can be created in the place Matrix, which is described in the next paragraph; 2) the degree of longitude and latitude as decimal, and 3) the capacity of each customer.

To this end the public class placeMatrix is called in the main class. Here a matrix gets filled with the values of the degree of latitude and longitude and the capacity of each customer. The matrix will have as many rows as existing customers and four columns. In the first column the algorithm counts from 1 to i in order to better distinguish between customers. The second column contains the degree of latitude and the third one the degree of longitude, both as a decimal.

In the last column is the value of the capacity. Once all the data from the customers are entered the class returns the matrix to the main class.

The algorithm searches in the first for-loop for the lowest and highest value of the degree of latitude and longitude. If the latitude of customer i is lower than the current one then the latitude of customer i is now the lowest one. However, if the latitude of customer i is higher than the current one then the latitude of customer i will be the highest one. This operation repeats until the algorithm goes through all the values in the second column. For the degree of longitude, the algorithm finds the lowest and highest value in the same way in the third column of the matrix. This is necessary in order to narrow the area to a rectangle in which the optimal solution is found and to shorten the runtime of the algorithm.

The algorithm further calculates the Euclidean distance from every customer considering its possible demand on each of the potential hub locations. Therefore, two while-loops are implemented. The first one counts up from the shortest distance based on latitude point to the largest one and the second loop counts from the shortest distance based on longitude point to the largest one. So, the algorithm starts calculating with the smallest degree value of each as hub location with the coordinates (smallest x/ smallest y). The first iteration calculates the Euclidian distance from the hub when it is built at the bottom left corner of the rectangle to every customer. The code below illustrates a section of the procedure.

```
while(shortestX < largestX) { /*As long as shortestX is smaller than the largest x and*/
    while(shortestY < largestY) { /*as long as shortestY is smaller than the largest y*/
        for(int i = 0; i < p.length; i++) { /*the distance from customer i to the potential hub with the coordinates(shortestX/shortestY) is calculated*/
            double a = Math.pow((p[i][1] - shortestX),2); /*by calculating the square of the latitude of location i minus the latitude of the potential hub*/
            double b = Math.pow((p[i][2] - shortestY), 2); /*and by calculating the square of the longitude of location i minus the longitude of the potential hub*/
            shortestD=shortestD+(Math.pow((a+b),0.5) *p[i][3]); /*the distance is now calculated by the root extraction of the addition from the first two results multiplied with the capacity of i*/

            if(shortestDOld == 0) { //if this is the first iteration
                x = shortestX; //x is set as the first optimal solution
                y = shortestY; //y is set as the first optimal solution
                shortestDOld = shortestD;
            }
            else if(shortestD < shortestDOld) { /*after the first iteration we only set x and y when the new distance is shorter than the old one*/
                shortestDOld = shortestD;
                x = shortestX;
                y = shortestY;
            }
            shortestY = shortestY + 0.00001; /*the degree of latitude of the potential hub is changed and the iteration starts again from while(shortestY < largestY)*/
            shortestD = 0; /*shortest D has to be set back to calculate a new distance*/
        }
        shortestY = yOld; /*if shortestY is not smaller than the largest y anymore, y has to be set back to the initial value*/
        shortestX = shortestX + 0.00001; /*and the degree of longitude of the potential hub is changed and the whole iteration starts again until shortestX is no longer shorter than the largest x and all distances have been calculated*/
    }
    Out.print("\n\nHub should be built on the degree of latitude " + x + " and on the degree of longitude " + y);
}
```

The Euclidean distance is denoted as follows: $\Sigma$ over all i $\sqrt{((\text{latitude of customer i} - \text{latitude of hub location})^2 + (\text{longitude of customer i} - \text{longitude of hub location})^2)} * $ capacity of the customer i. The variables x, y and shortestDOld get set with the first results in the first iteration. In the next iterations the variables are only set if the new distance is shorter than the old one. At the end of the second while-loop the shortestY value is increased by 0.00001.

This cycle is then repeated until shortestY is as large as largestY and shortestX gets increased by a value of 0.00001. ShortestY is set back to its original value and the whole cycle starts again until shortestX is as large as largestX. After this the algorithm determines the optimal location for the hub.

## V. ROUTING ALGORITHM

We assume that packages have the dimensions 20x40x30cm. The duration and visual representation of the route was calculated as an estimation according to the values in the ÖAMTC Austrian automobile association [9]. The route can differ according to traffic jams, road construction, accidents, etc.

To calculate the optimal route, we selected the A* algorithm [10] because of its better run-time in contrast to for example the Dijkstra algorithm [11]. Instead of calculating every possible solution as the Dijkstra algorithm does, the A* algorithm estimates whether a solution could be the optimal one or not and only calculates the paths that could lead to an optimum. Moreover, the algorithm also considers obstacles within the route, like toll fees along the route. However, in order to work for this model, the A* algorithm had to be adopted in some parts of the main class of the hub location algorithm that was described in section IV.

To adopt the routing algorithm in our approach we had to first initialize each potential customer in a separate node with the variable type (String, double). In this case the "String" value is the street name and the "length" value is the distance from the hub to the customer in kilometers. We show an example below.

//initialize the graph based on the map of Linz

Node n1 = new Node("Europastraße",3);

Moreover, each node has to be connected to the other customers and also to the hub again. Since Linz has many one-way streets in some cases the distance in one direction is longer than in the other one. As shown in the code below one edge is defined as Node, double. Here the target node is needed to which the current initialized node should be connected. Furthermore, for the double variable the distance between the current one to the target node is used.

//initialize the edges

//Europastraße

n1.adjacencies = new Edge[]{

   new Edge(n2,0.8)}

At the end of the main class, the A* algorithm calls up the method AstarSearch (Node, Node) where it calculates the shortest path between the inserted two nodes. As an example, we show the path from node 17, "Fadingerplatz", and node 13, the hub, searched and displayed.

AstarSearch(n17,n13);

   List<Node> path = printPath(n13);

      System.out.println("Path: " + path);

On the console it is displayed as follows:

Path: [Fadingerplatz, Wiener Straße 226, Muldenstraße 2, Bulgariplatz, Hub]

### A. Hub Location

Table I shows the locations at which the hubs should be built for each zone as a result of applying the algorithm presented in the previous section. It depicts the coordinates of the hub as well as the address and the average time and distance which is needed to travel from the hub to each customer.

To validate the results of the hub location algorithm, we measured the joint variability of the variables minutes and kilometers for all the zones. As depicted in Fig. 2 the measures showed a positive covariance. Table 2 shows the strength of the relationship between the resulting kilometers and minutes in Zones 1, 2 and 3. As it can be seen the correlation coefficients show a strong linear relationship.

**Zone 1**: The resulting location is near an arterial road with two lanes each way and it is close both to the access road to the Highway A7 and to the exit Linz-Urfahr. However, the problem is that at this location there is a chain furniture store and a shoe shop. To locate the hub at these exact coordinates, the stores would have to move to a different location. Further, according to the storage capacity previously defined in section IIA, which considered the existing pick-up stations and parcel lockers from the post, the hub would be required to have a capacity of 595 packages to be able to satisfy the largest possible demand. In this work, we assume that the hub can be built at the exact address that resulted from applying the algorithm and that the construction of a space with the maximum storage capacity is guaranteed.

TABLE I. LOCATIONS AT WHICH THE HUBS SHOULD BE BUILT FOR EACH ZONE, DISTANCES AND TIME TO TRAVEL FROM HUB TO CUSTOMER. THE RESULTING ADDRESSES ARE AS FOLLOWS: 1: FREISTÄDTERSTR. 91-93L, ADDRESS 2: SÜDTIROLERSTRAßE 8 AND ADDRESS 3: WANKMÜLLERHOFSTRAßE 34

| Hub Location | Average Distance | | | | | STD | |
|---|---|---|---|---|---|---|---|
| | $\varphi$ | $\lambda$ | Address | Min | Km | Min | Km |
| Zone 1 | 48.321556 | 14.290709 | 1 | 6,63 | 2,54 | 3,43 | 1,67 |
| Zone 2 | 48.298063 | 14.292985 | 2 | 6,85 | 2,17 | 3,15 | 2,17 |
| Zone 3 | 48.28098 | 14.3029 | 3 | 5,41 | 3,1 | 2,31 | 1,58 |

TABLE II. STRENGTH OF THE RELATIONSHIP BETWEEN THE RESULTING KILOMETERS AND MINUTES

| Hub Location | Correlation Coefficient r |
|---|---|
| Zone 1 | 0,87 |
| Zone 2 | 0,81 |
| Zone 3 | 0,93 |

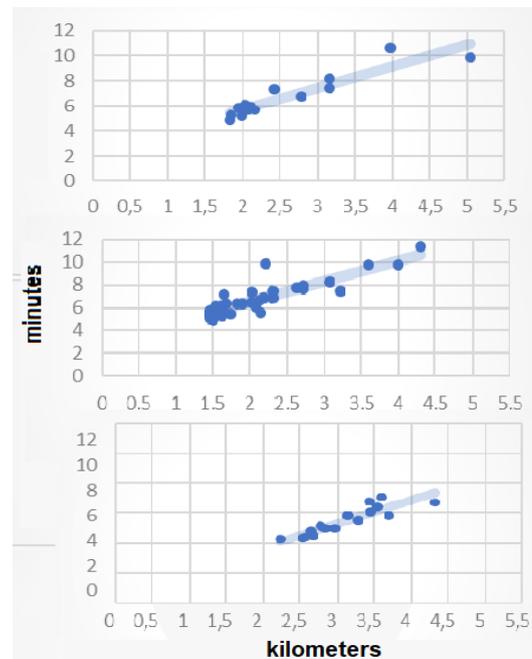

Figure 2. Positive covariance measured from the joint variability of the variable's minutes and kilometers in zone 1, 2 and 3 (top to bottom).

**Zone 2**: Even though the hub is quite near to the customers, this location in zone 2 is not optimal in reality. Since this place is located in Linz downtown, trucks would have to drive a long distance from the highway to the hub. Moreover, the streets in the inner city are all quite narrow and most of them are one-way streets.

Furthermore, this hub would need a capacity for 1405 packages, which requires an amount of space not available in the inner city.

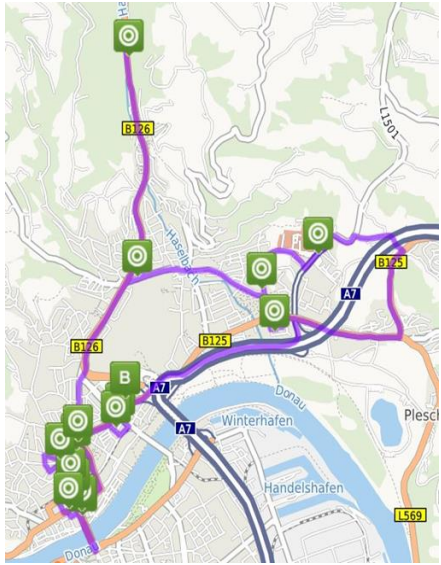

Figure 3. Route for zone 1 resulting from the A*Algorithm

However, to continue with the model, it is assumed that it is possible to build the hub at this exact place and with this capacity.

**Zone 3**: The resulting coordinates for the hub in zone 3 (48.265266 lat., 14.2975669 long.) are at the highway Mühlkreisautobahn around the exit Wiener Straße. In order for this model to be continued, it is therefore assumed that the optimal location for the hub is at the degree of latitude 48.28098 and at the degree of longitude 14.3029, as depicted in Table I, which would be the address Wankmüllerhofstraße 34. This place is acceptably near to the original optimal solution. On average it takes 3,1 kilometers to travel from the hub to the customer.

An advantage for this location is its proximity to the freeway exit Wiener Straße and therefore truck drivers do not have to drive through the smaller city streets. This hub would need a capacity of 945 packages to satisfy the highest assumed demand.

### B. Routing Algorithm

Regarding the results of the A* algorithm, in the graphics showed in this section the points "B" mark the hub locations and the green squares with the circles inside represent the customer locations. The use of kilometers or minutes in the algorithm did not affect the route solution.

**Zone 1**: the best route for Zone 1 is represented in Fig. 3. This route should take with an estimated stopping time of 7 minutes (including stopping, delivering the packages and starting the truck again) approximately 160 minutes. The highest possible supposed demand is 595 packages. Hence, either a semi-trailer is used, which would not be as easy to park and manoeuvre in the city, or the route is split into two separate routes and two big transporters with a load capacity of around 20 cubic meters.

In this case route 1 would cover the customers placed north-east of the hub. It would take 73 minutes including the estimated 7 minutes stopping time and a maximum demand of 290 packages, the path being as follows: Hub – Leonfeldnerstraße 328 – Ferdinand-Markl-Straße 5 – Johann-Wilhelm-Klein Straße 23 – Aubrunnerweg 1 – Mostnystraße 5 – Hub

Route 2 would take about 97 minutes including the 7 minutes stopping time and delivering a maximum of 305 packages: Hub – Linke Brückenstraße 15 – Linke Brückenstraße 13 – Hauptstraße 10 – Friedrichstraße 2 – Hauptstraße 30 – Hauptstraße 46 – Rudolfstraße 14 – Nestroystraße 2 – Wischerstraße 1 – Freistädter Straße 29 – Hub

**Zone 2**: Fig. 4 represents the path for Zone 2. The stopping time for this route is also 7 minutes and therefore it takes about 339 minutes to deliver a maximum of 1405 packages. Since all packages would not fit in a semi-trailer and a driving time of more than 5 and a half hours at once is not desired, the route has to be split into 3 different routes.

Route 1: with a maximum capacity of 455 packages and 111 minutes driving time, including the 7 minutes stopping time at each customer: Hub – Bürgerstraße 14 – Bismarck 9 – Fadingerstraße 11 – Landstraße 17-25 – Domgasse 1 – Graben 19a – Obere Donaulände 1 – Lessingstraße 9 – Altstadt 12 – Herrenstraße 27 – Herrenstraße 46 – Hub

Route 2: 127 minutes delivery time for a maximum of 500 packages: Hub – Lüfteneggerstraße 10 – Industriezeile 78 – Ignaz-Mayer-Straße 7 – Köglstraße 22 – Köglstraße 19 – Franckstraße 15 – Lonstorferplatz 1 – Hamleringstraße 44 – Goehtestraße 52 – Goethestraße 41 – Scharitzerstraße 23 – Goethestraße 22 –Hub

Route 3: with a delivery time of 94 minutes including 7 minutes to stop for a maximum of 450 packages: Hub – Volksgartenstraße 1 – Stockhofstraße 9 – Bahnhofplatz 11 – Waldeggstraße 44a – Bockgasse 30 – Unionstraße 100 – Unionstraße 75 – Unionstraße 5 – Schillerstraße 4 – Hub

**Zone 3**: Fig. 5 represents the path for Zone 3. This tour will take, including an estimated stopping time of 7 minutes, about 153 minutes. The maximum total demand of all customers is 795 packages. Depending on the size of the different packages they probably will not fit all together in one truck. Hence, the route has to be split in two separate routes.

Route 1: 365 packages will be delivered with the estimated stopping time of 7 minutes in about 86 minutes: Hub – Landwiedstraße 65 – Europastraße 12 – Einsteinstraße 5 – Muldenstraße 33 – Muldenstraße 2 – Wiener Straße 93 – Wiener Straße 99 – Gürtelstraße 5 – Bulgariplatz 1 – Hub.

Route 2: The maximum total demand of these customers is 430 which also should fit in one semi-trailer. As well as in the route before the stopping time is assumed to be 7 minutes. So, the delivery time should be about 77 minutes: Hub – Wiener Straße 226 – Denkstraße 37 – Dauphinestraße 28 – Wiener Straße 437 – Fadingerplatz 6 – Dieselstraße 4 – Benzstraße 14 – Hub

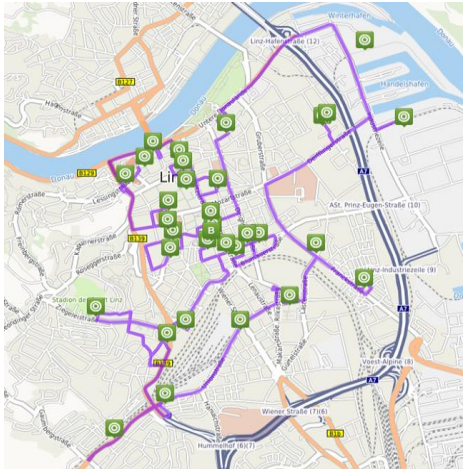

Figure 4. Route for zone 2 resulting from the A*Algorithm.

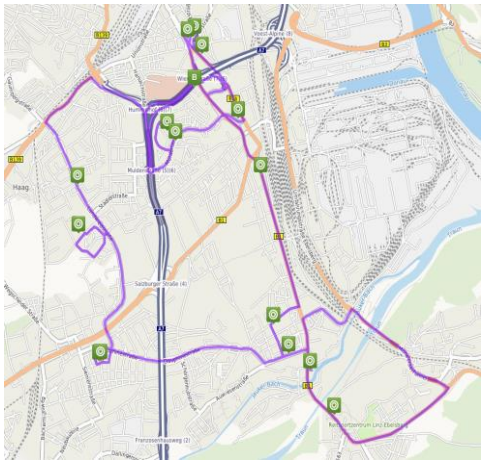

Figure 5. Route for zone 3 resulting from the A*Algorithm

## VI. CONCLUSION AND FUTURE WORK

Reducing CO2 emissions is a pressing measure in order to slow/reduce climate change. To this end it is essential that companies cooperate to use less resources, including fuel and vehicles, and that common mandatory regulations in the European Union exist. In an urban environment, the implementation of hubs could alleviate traffic. The model presented in this work focuses on Linz, Austria, but it could be transferred to other cities as well. The presented approach tackles the last 2 to 5 kilometres of delivery that are in close proximity to the customer. It reduces driving time and distances, resulting in an increase of short haul trips that also benefit the employee drivers, in that they also have a shorter commute to work. The three created hubs will make it possible for the trucks to drive directly to their destination without having to cross the city. In the routes to the hubs, the traffic might increase, but to deliver the packages to all of the pick-up stations and parcel lockers only 7 big transporters are needed. Hence, the overall traffic within the city would benefit from the presented measure in this paper. Future work will focus on extending the approach to other cities.

ACKNOWLEDGMENT

This work was supported by the Austrian Ministry for Climate Action, Environment, Energy, Mobility, Innovation and Technology (BMK) Endowed Professorship for Sustainable Transport Logistics 4.0.